\documentclass[mathleft]{an}
\usepackage{graphicx}
\usepackage{times}
\usepackage{natbib}
\bibpunct{(}{)}{;}{a}{}{,}
\newcommand{\edit}[1]{\textbf{#1}}
\renewcommand{\edit}[1]{#1}

\begin{document}


\Pagespan{789}{}
\Yearpublication{2013}%
\Yearsubmission{2012}%
\Month{11}%
\Volume{999}%
\Issue{88}%


\title{%
  Stability of librational motion in the spatial circular restricted three-body problem for high inclinations and mass ratios
}

\titlerunning{%
  Stability of librational motion in the CR3BP
}

\author{%
  {\'A}. Bazs{\'o}\inst{1}\fnmsep\thanks{Corresponding author: \email{akos.bazso@univie.ac.at}\newline}
  \and
  R. Schwarz\inst{1}
  \and
  B. {\'E}rdi\inst{2}
  \and
  B. Funk\inst{1}
}

\authorrunning{%
  Bazs{\'o}, Schwarz, {\'E}rdi \& Funk
}

\institute{%
  Institute for Astrophysics, University of Vienna,
  T{\"u}rkenschanzstra{\ss}e 17, A-1180 Vienna, Austria
  \and
  Department of Astronomy, E{\"o}tv{\"o}s University,
  P{\'a}zm{\'a}ny P{\'e}ter s{\'e}t{\'a}ny 1/A, H-1117 Budapest, Hungary
}

\received{XXX}
\accepted{YYY}
\publonline{later}



\abstract{%
In the spatial circular restricted three-body problem librational motion around the Lagrangian points $L_4$ and $L_5$ exists up to high inclinations of the massless object. We report the results of numerical investigations on the stability of this librational motion for systematic variations in the inclination of the Trojan and the mass-ratio $\mu$ of the two massive bodies. We show that stable motion prevails for inclinations below 60 degrees and mass-ratios $\mu < 0.04$ for typical integration times up to $10^6$ revolution periods. At even higher mass-ratios -- beyond the critical mass-ratio for the planar case -- stable orbits were found to exist for up to $10^7$ periods at moderate inclinations. We extracted the librational frequencies on a grid in the parameter space from the Fourier spectra and traced their variation. Several resonances between the short and long period librational frequency as well as the vertical frequency lie inside the investigated region. The application of the Lyapunov characteristic indicator and spectral \edit{number} methods also reveals the chaotic regions. This simple model is \edit{equally} applicable in the Solar system for low mass ratios for Trojans of the planets, as well as to Trojan-type exoplanets in binary star systems at high mass-ratios.
}

\maketitle


\section{Introduction}

The circular restricted three-body problem (CR3BP) seems to be a simplistic model of not much practical importance. But in fact it serves as a reasonable first approximation in many cases, for instance in the Solar system or in extrasolar planetary systems.

\subsection{Solar system}

In the Solar system case, the investigation of the dynamics of co-orbital asteroids moving in a 1:1 mean-motion resonance with a planet, the ``Trojan'' asteroids, is the goal ever since the problem has been posed by Euler and Lagrange in the 18th and 19th century. Currently Trojan asteroids have been found for the planets Earth, Mars, Jupiter, and Neptune, see e.g. \citet{Wolf1907,She2006,Con2011}. Interestingly no Trojans were detected for Saturn and Uranus despite serious efforts \citep{Mar2002}. After the discovery of Neptune Trojans with inclinations in excess of $20^\circ$ interest arose in the dynamics of objects at even higher inclination. For Neptune \citet{Zho2009} determined three regions of stable motion, one being at inclinations larger than $50^\circ$. For Uranus a similar study by \citet{Dvo2010} concluded that Trojans on low inclination orbits could survive over timescales of $10^9$ years. Both studies used as dynamical model the outer Solar system including all four gas planets, thus accounting for mutual perturbations of the planets; but a more complex model can also obscure the fundamental features of Trojan motion. As the planets' eccentricities are small enough in the Solar system, it is justified to apply the CR3BP to determine the basic stability regions \citep{San2002,Eft2005} before advancing to more sophisticated models like the elliptic restricted three-body problem or the N-body problem.


\subsection{Extrasolar systems}

For the extrasolar case theoretical interest in Trojan-type (T-type) motion began even before the discovery of the first extrasolar planet. In \citet{Rab1988} different types of planetary motion in double star systems were considered. Besides the planetary (P-type) and satellite (S-type) motion around both respectively one of the components, T-type motion was considered as a third possibility. Not only the formation of Earth-mass planets directly in T-type orbits (about a migrating gas giant in a proto-planetary disk) was studied \citep{Bea2007}, but also the possible capture of asteroids into Trojan orbits by the inner planets \citep{Sch2012a}. The stability regions of T-type motion close to the habitable zone in specific extrasolar planetary systems were investigated by \citet{Dvo2004,Erd2005,Sch2009}, and additionally in \citet{Fun2012}. Since extrasolar planets were also detected in binary star systems and systems consisting of a star and a sub-stellar companion, the dynamics of librational motion at high mass ratios is of interest, especially for the non-coplanar motion.


\section{Model \& methods}

The CR3BP consists of two massive objects ($m_1$, $m_2$), called primaries, and one object of negligible mass (considered to be massless) relative to the others. The primaries move on circular orbits about their common barycenter;
\edit{while the third body (the ``Trojan'') performs a three dimensional librational motion in the vicinity of the Lagrangian equilibrium points $L_4$ or $L_5$.}

For the CR3BP it has been shown by \citet{Gas1843} and \citet{Rou1875} that for mass ratios in the interval $0 \le \mu < \mu_c$ the librational motion is linearly stable (here $\mu = m_2/(m_1+m_2)$, and $\mu_c = ( 1 - \sqrt{23/27} ) / 2 \approx 0.0385$). For the non-linear system the same is true except for three special values of $\mu$ that correspond to resonances of the librational frequencies \citep{Mar1972,Whi1983}.

Usually the dimensionless masses are normalised to $m_1 + m_2 = 1$ which results in $\mu_{c} = 0.0385$ as above, but following the article of \citet{Erd2007} here we use a different normalisation: $m_1 + m_2 = 1 + \mu$ giving $\mu_{c} = 0.04$. The conversion from the first to the second normalisation is achieved by multiplying all mass ratios with $1 + \mu$.

On a grid of $100 \times 61$ initial conditions we vary the mass ratio in the range $0.0005 \le \mu \le 0.05$ and the inclination $0 \le i \le 60$ degrees, respectively. These ranges result in a coverage of mass ratios from approximately Jupiter mass planets up to sub-stellar and stellar companions, when the mass $m_1$ is assigned one solar mass. The Trojan is always started in the vicinity of the Lagrangian point $L_4$. The equations of motion were integrated numerically, two methods were applied: (i) a Lie integrator with adaptive step-size \citep{Han1984,Egg2010}, and (ii) a Bulirsch-Stoer integrator that also handles the variational equations for an estimate of the Lyapunov characteristic exponent.

In a first run the integration time was set to $10^6$ orbital periods to determine regular and chaotic domains inside the parameter space. From this run we obtained the maximum eccentricity (ME), libration amplitude (LA), and Lyapunov characteristic indicator (LCI), a finite time estimate of the maximum Lyapunov exponent. For the long term integrations ($10^7$ orbital periods) we focused on the high mass ratio border and investigated the interval $0.0385 \le \mu \le 0.046$ for inclinations $30^{\circ} \le i \le 50^{\circ}$. Finally, the frequencies of librational motion were determined through short time integrations of $10^3$ periods on the grid of initial conditions mentioned above. \edit{In some cases a longer integration time did not allow to determine the librational frequencies with enough accuracy}.

The Laplace-Lagrange variables $h = e \sin (\omega + \Omega)$ and $p = \sin i \sin \Omega$ were analysed with the Discrete Fourier Transform \citep{Ree2007} and Fast Fourier Transform techniques\footnote{FFTW: M. Frigo \& S.~G. Johnson, http://fftw.org}. The sampling interval was chosen to be 18 days, which gives just more than $2\times10^4$ samples over 1000 periods.


\section{Results}

\subsection{Spectral number}

\begin{figure}
{%
\includegraphics[width=0.48\textwidth]{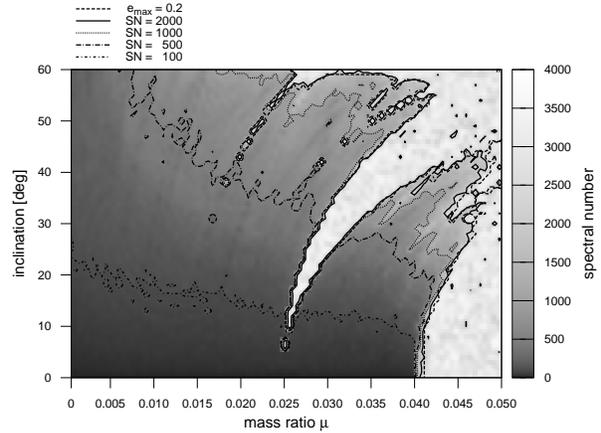}
\caption{A spectral number (SN) map of the CR3BP depending on mass ratio and inclination. \edit{The gray-levels represent the number of peaks in the power spectra of individual initial conditions; a higher number means more pronounced chaotic behaviour}. As an overlay contour lines for different SN values are plotted; \edit{additionally the $e_{\mathrm{max}}=0.2$ contour line from the $10^6$ period integration indicates the border of stable librational motion. The contour labels correspond to the contour lines from top to bottom.}
}
\label{fig:specindex}
}
\end{figure}

From the Fourier analysis of each short integration the number of peaks in the obtained power spectrum was counted giving a measure called ``spectral number'' (SN). The spectral number method has been applied e.g. in \citet{Mic1995,Zho2009} for a similar purpose. A low number of frequencies in the power spectrum indicates a rather regular orbit, while a large number of peaks is evidence for a highly chaotic orbit. We did not apply a low pass filter, which would remove many tentative peaks, but still the resulting Figure \ref{fig:specindex} shows clearly the border between regular and chaotic regions. For the planar case and at low inclination we find regular motion with only a few dozen peaks. As expected from the linear stability analysis the transition from regular to chaotic motion in the planar case occurs at $\mu = 0.04$, while at higher inclinations the border shifts to mass ratios $\mu > \mu_c$. In \citet{Sic2010} it has already been indicated that also for the planar case stable periodic orbits exist for $\mu > \mu_c$. At high inclinations the number of peaks increases; however there is a clear discontinuity regarding the number of peaks between the regions of regular and chaotic motion.

\subsection{Maximum eccentricity \& libration amplitude}

The SN does not necessarily give an indication about long-term stability, for this purpose the ME was considered. The results for the long term integrations ($10^6$ and $10^7$ periods) are not shown here (see \citet{Sch2012b} for details), \edit{but in summary the ME gives almost exactly the same shape for the regular region as the SN in Figure \ref{fig:specindex}}. If the maximum eccentricity remains $e \le 0.2$ then the librational motion lasts up to $10^7$ periods of the primaries.
\edit{At the ragged right edge at high mass ratios a small discrepancy between SN and ME is visible in Figure \ref{fig:specindex}. These are chaotic orbits with high SN that stay with low ME for long times, we attribute this to the ``stickiness phenomenon'' \citep{Dvo1998}.}

Using the libration amplitude $\sigma = \lambda_T - \lambda_S$, which is the difference of the mean longitudes ($\lambda = \omega + \Omega + M$) for the Trojan and the less massive primary, we check for a permanent librational motion. If this amplitude exceeds $180^{\circ}$ the Trojan would not librate about the Lagrange point but rather be in a horseshoe orbit. The results show small LA for low initial inclinations, and the amplitudes smoothly increase with inclination. The largest LA inside the regular domain are $\sigma < 30^{\circ}$, confirming the results obtained with the SN and ME.


\section{Discussion}

\subsection{Libration frequencies}

In a recent investigation on the elliptic restricted three-body problem (ER3BP) for high inclinations \citet{Sch2012b} found that secondary resonances between the libration frequencies \citep{Erd2007,Erd2009} play a crucial role for the long-term stability. We then reviewed the CR3BP and were searching for similar effects. As \citet{Sze1967} points out, the linear variational equations of the spatial CR3BP can be separated into the planar motion, involving the short and long period components $n_s$ and $n_l$, which is decoupled from the vertical motion. This vertical motion -- in the linearized case -- is simply a harmonic oscillator with frequency $n_z = \sqrt{ -\Omega_{zz}(L_{4,5}) }$, where $\Omega_{zz} = \partial^2 \Omega / \partial z^2$ is the second partial derivative of the potential function in the rotating coordinate system. Exactly at the Lagrangian points $n_z = 1$, which is equal to the dimensionless orbital frequency (mean motion) of the primaries. Taking into account higher order terms there is a non-linear coupling between the $\{ x,y,z \}$-components of the equations of motion.

In the spatial case \citet{Bra2004} have shown for a limited range of mass-ratios that \edit{the region for} librational motion ceases at an inclination of $i = 61.5^\circ$ for planetary mass objects, and is linearly decreasing with increasing mass. They argue that this limit is caused by saddle-points in the averaged perturbing potential function.

\begin{figure}
{%
\includegraphics[width=0.45\textwidth]{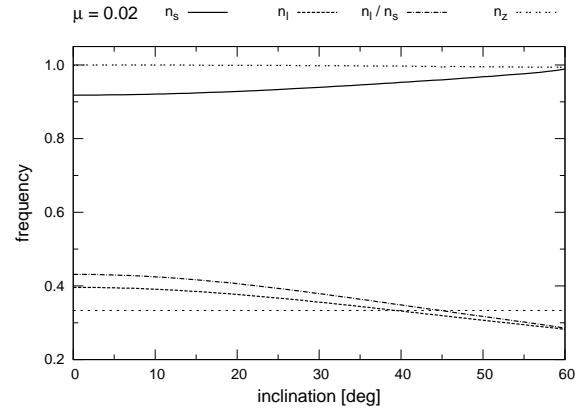}
\caption{The variation of librational frequencies $n_s$ and $n_l$ depending on the inclination for the mass ratio $\mu = 0.02$. The top dotted line is the vertical frequency $n_z$, $n_s$ is increasing while $n_l$ is decreasing. \edit{At $i \approx 45^{\circ}$ $n_s = 3 n_l$ leading to a resonance (B 3:1), while} at $i \approx 60^{\circ}$ $n_z$ and $n_s$ become comparable, resulting in a 1:1 resonance between vertical and short period librational frequency.}
\label{fig:libfreq1}
}
\end{figure}

\begin{figure}
{%
\includegraphics[width=0.45\textwidth]{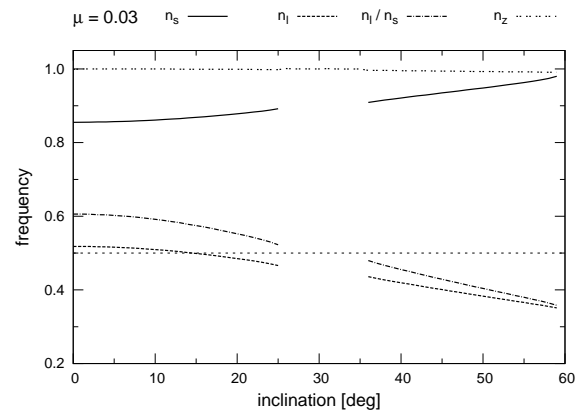}
\caption{The variation of librational frequencies with inclination for $\mu = 0.03$. Around $i = 30^{\circ}$ the frequencies $n_l$ and $n_s$ (their ratio is plotted as the dash-dotted line) are in a 1:2 resonance (indicated by the horizontal dotted line). This resonance is causing strongly chaotic motion and libration is not possible. The gap in the data means that there the frequencies could not be determined, as these initial conditions lie inside the chaotic zone, compare Figure \ref{fig:specindex}.}
\label{fig:libfreq2}
}
\end{figure}

From Figure \ref{fig:libfreq1} it is visible that the libration frequencies $n_s$ and $n_l$ vary with the inclination, while the vertical frequency $n_z$ changes only negligibly. The short period component $n_s$ tends to unity, while the other component $n_l$ is monotonically decreasing. As $i$ approaches $60^{\circ}$ (depending on the mass ratio) the frequencies $n_s$ and $n_z$ become commensurable, i.e., they are in a 1:1 resonance. Thus the high inclination border around $60^{\circ}$ can be explained through Figure \ref{fig:libfreq1}.
\edit{Another feature is the $n_s : n_l =$ 3:1 resonance crossing around $45^{\circ}$ (i.e., when crossing the horizontal line, cf. Figure \ref{fig:specindex} for $\mu = 0.02$), but contrary to Figure \ref{fig:libfreq2} no gap is visible, because the librational frequencies are well defined despite the high SN.}

Another example is shown in Figure \ref{fig:libfreq2}, for a resonance between $n_s$ and $n_l$. At the chosen mass ratio of $\mu = 0.03$ around an inclination of $i = 30^{\circ}$ the long-period and short-period components are commensurable: $n_s : n_l =$ 2:1. This resonance leads to strongly chaotic motion within $10^3$ periods of the primaries and causes the big gap in Figure \ref{fig:libfreq2}, which is also visible in Figure \ref{fig:specindex}.


\subsection{Lyapunov indicator}

\begin{figure}
{%
\includegraphics[width=0.45\textwidth]{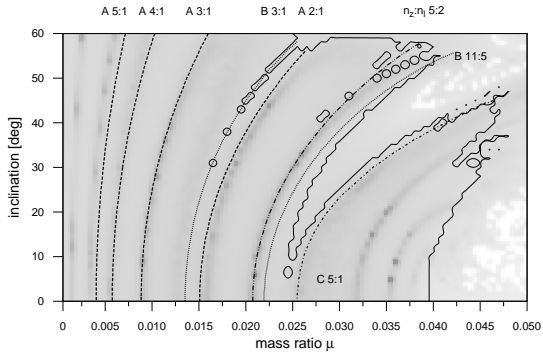}
\caption{The Lyapunov characteristic indicator after 100 periods is assigned as gray values to the grid of $\mu-i$. For visualising the border of regular motion the $e_{\mathrm{max}} = 0.2$ contour is plotted. The various lines represent the resonances between the librational frequencies.}
\label{fig:lci}
}
\end{figure}

In the ER3BP the short and long period libration frequencies $n_s$ and $n_l$ \edit{can form} combinations with the elliptic motion of $L_4$ with \edit{normalised} frequency $n = 1$ and give rise to different secondary resonances \citep{Erd2007}. In the CR3BP there is no elliptical motion of the Lagrange point itself, \edit{but nevertheless the frequencies can form combinations with the vertical frequency $n_z$ (instead of $n$), like $n_z - n_s$, $n_z - n_l$.} Following the nomenclature of \citet{Erd2007,Erd2009} we assign letters to the combinations, where e.g. $A = (n_z - n_l) / n_l$, $B = n_s / n_l$, $C = (n_z - n_l) / (n_z - n_s)$, \edit{keeping in mind that $n_z = 1$ at $L_4$}. Calculating the locations of resonances from the grid of $\{ n_s, n_l \}$ we obtain Figure \ref{fig:lci}.

For the high mass ratio border the {B 1:1} resonance limits the librational motion originating from the point $(\mu_c,0)$ \citep{Dan1964,Erd2007}. A number of type A resonances appear in the low mass region, while the B 2:1 resonance (see Figure \ref{fig:libfreq2}) is responsible for the large gap. It is surrounded by various other resonances like the B 11:5 and C 5:1, where the latter starts close to $(\mu,i)=(0.025,0)$ and follows closely the border of the gap. Also visible in Figure \ref{fig:lci} is the $n_z/n_l=5/2$ resonance that occurs whenever $n_l \approx 0.4$, originating between $0.020 < \mu < 0.021$.


\section{Summary}

We find for the CR3BP regular librational motion for mass ratios up to and above the critical mass ratio $\mu_c$ for inclined orbits. Librations around $L_4$ are stable for $10^{6}-10^{7}$ periods also at high inclinations (up to the considered limit of $i = 60^{\circ}$), and for high mass ratios. The application of different methods (LA, LCI, ME, SN) confirms that the structures found are truly related to the librational motion. \edit{The frequencies $n_s, n_l, n_z$ can form combinations, when they appear as an integer ratio they cause a resonance leading to chaotic motion. Several such resonances have been identified, mainly the $A$ and $B$ types together with such involving the vertical frequency $n_z$}. At inclinations of about $i=60^{\circ}$ (depending on the mass ratio) the frequencies $n_s$ and $n_z$ become equal and then librational motion is not possible any more. These results on the CR3BP can be applied to Trojan type motion in the Solar system, and especially in the high mass ratio region to extrasolar star--brown-dwarf and binary-star systems that could harbour Trojan planets.


\acknowledgements
{\'A}. Bazs{\'o} acknowledges the support from the Austrian FWF project P23810-N16 and the doctoral school ``Planetology: From Asteroids to Impact Craters'' at the University of Vienna.
R. Schwarz wants to acknowledge the support by the Austrian FWF project P23810-N16.
B. Funk wants to acknowledge the support by the Austrian FWF projects P22603-N16 and P23810-N16.



\end{document}